\documentclass{doublecol-new}
\usepackage{natbib}
\usepackage{mathrsfs}
\usepackage{wrapfig} 
\usepackage{url}
\usepackage{hyperref}
\hypersetup{
    colorlinks,
    linkcolor={red!40!black},
    citecolor={blue!40!black},
    urlcolor={blue!70!black}
}
\usepackage{booktabs}

\usepackage{microtype}
\usepackage{placeins}
\usepackage{epstopdf}
\usepackage{multirow}%
\usepackage{optidef} % 
\usepackage{amsmath,amssymb,amsfonts}%

\usepackage{textcomp}%
\usepackage{manyfoot}%
\usepackage{booktabs}%
\usepackage{algorithm}%
\usepackage{algorithmicx}%
\usepackage{algpseudocode}%
\usepackage{listings}%
\usepackage{optidef}
\usepackage[capitalize]{cleveref}
\usepackage{mathtools} 
\usepackage{verbatim}
\usepackage{graphicx}
\usepackage{subfigure}
\usepackage{caption}

\usepackage{tabularx}
\usepackage[export]{adjustbox}
\usepackage{xcolor}
\usepackage{enumitem}

\hypersetup{colorlinks, linkcolor={red!40!black}, citecolor={blue!40!black}, urlcolor={blue!70!black}}

\theoremstyle{TH}{

}

\theoremstyle{THrm}{

}

\theoremstyle{THhit}{

}
        
\makeatletter

\def\theequation{\arabic{equation}}

\makeatother

\begin{document}

\setcounter{page}{1}

\LRH{}

\RRH{}

\VOL{x}

\ISSUE{x}

\PUBYEAR{xxxx}

\PUBYEAR{201X}

\subtitle{}

\title{Enhancing Group Recommendation using Soft Impute Singular Value Decomposition}
\authorA{Mubaraka Sani Ibrahim}
\affA{Department of Computer Science,\\ African University of Science and Technology,\\ Abuja, Nigeria \\
msani@aust.edu.ng}

\authorB{Isah Charles Saidu}
\affB{Department of Computer Science,\\ Baze University,\\ Abuja, Nigeria \\
charles.isah@bazeuniversity.edu.ng}
\authorC{Lehel Csat\'o}
\affC{Faculty of Mathematics and Computer Science,\\ Babes-Bolyai University,\\ Cluj-Napoca, Romania \\
lehel.csato@ubbcluj.ro}

\begin{abstract}
The growing popularity of group activities increased the need to develop methods for providing recommendations to a group of users based on the collective preferences of the group members. Several group recommender systems have been proposed, but these methods often struggle due to sparsity and high-dimensionality of the available data, common in many real-world applications. In this paper, we propose a group recommender system called Group Soft-Impute SVD, which leverages soft-impute singular value decomposition to enhance group recommendations. This approach addresses the challenge of sparse high-dimensional data using low-rank matrix completion. We compared the performance of Group Soft-Impute SVD with Group MF based approaches and found that our method outperforms the baselines in recall for small user groups while achieving comparable results across all group sizes when tasked on Goodbooks, Movielens, and Synthetic datasets. Furthermore, our method recovers lower matrix ranks than the baselines, demonstrating its effectiveness in handling high-dimensional data.
\end{abstract}

\KEYWORD{collaborative filtering; matrix completion; matrix factorisation; nuclear norm; rank; recommendation system; sparsity.}

\maketitle

\section{Introduction}
Recommendation system is an algorithm that provides personalized recommendations to users based on preferences, past behaviour, and similarities to other users. 

In general, there are two main types of recommendation systems: content-based filtering and collaborative filtering (CF). 
\noindent Content-based filtering \citep{b10} uses features to recommend items based on their similarity to items that the user has liked in the past. For instance,
when recommending books, various features of the book such as its publication year, genre, average rating, plot and characters are taken into account to create a recommendation list tailored to a user.

\noindent On the other hand, collaborative filtering \citep{b12} suggests items to users by analyzing the preferences of other users with similar interests. Collaborative filtering can be classified as memory-based CF and model-based CF:

\begin{enumerate}[label=\arabic*,left=0pt, itemindent=0pt, labelindent=0pt, labelsep=1.3em,itemsep=0.2em, align=left]
\item Memory-based CF aims to predict a target user's preferences for an item based on the preferences of similar users or a set of items \citep{b13}. Typically, similarity measures are used to identify a neighborhood of users (or items) that are similar to the target user (or item). Subsequently, the target user's interest in an item is determined based on the ratings of these neighboring users.
\item Model-based CF involves developing predictive machine learning models to forecast a user's rating for items they have not yet rated \citep{bib12}.
The model uses observed values in the user-item matrix as the training dataset and generates predictions for the missing values using the trained model. Model-based approaches include factorization methods, such as singular value decomposition (SVD) \citep{bib13}, Bayesian networks \citep{b14} and clustering algorithms \citep{b15}, such as k-means clustering. 
\end{enumerate} 

 A group recommender system generates personalized recommendations for a collection of users by aggregating individual preferences to formulate a suggestion that meets the needs of the group as a whole \citep{Jameson2007}. In contrast to traditional recommender systems, which focus on individual preferences, group recommenders must address the collective preferences, interactions, and potential conflicts within the group. These systems take into account various factors, including the diversity of preferences, the need for fairness in balancing different user inputs, and the dynamics within the group \citep{bib44}, such as majority rule, or consensus-based decision-making strategies.
 
In recent times, the popularity of group activities has increased the need to develop methods to cater to a group of users collectively, given their preferences. For instance, visiting a restaurant with friends, watching a movie with family, or trending content on social media websites are examples of activities well suited for groups of people.

There are several reasons for conducting group recommendation studies. First, most people engage in social activities with a group of people who are close acquaintances because humans are social by nature \citep{bib2}.  
Secondly, depending on technique, group recommendation may be more time efficient than personalized recommendation because estimating the individual preferences of users requires more computation time \citep{bib3}.
 
In recommender systems, users generally express their preferences for items by providing ratings. However, not all items receive ratings from users, resulting in a partially observed high-dimensional user-rating matrix with missing entries corresponding to unrated items. To address this, standard recommender systems aim to predict and impute these missing values, a process referred to as matrix completion \citep{bib22}.

In the context of group recommendation, we not only deal with a partially observed user-item matrix but also aim to generate recommendations at the group level, despite the sparse nature of the user-item interactions. This introduces an additional layer of complexity to the standard matrix completion problem, which is well-known to be NP-hard.

\noindent Two techniques commonly used to address group recommendation problems \citep{bib1}: aggregated predictions and aggregated models. The first approach involves generating preferences for each group member individually and then combining them to produce a group recommendation. In contrast, the second approach combines the preferences of individual group members into an extended dataset, which is subsequently utilised to generate recommendations for the entire group. Both methods rely on the chosen aggregation strategy.

In this study, we utilise the soft-impute algorithm \citep{bib33}, an efficient convex optimization technique for large-scale matrix completion. This approach operates under the assumption that group-level information is implicitly captured through the low-rank approximation of the partially observed user-item matrix. In our aggregated model-based approach, we leverage the extended dataset to ensure that items with more frequent ratings, as well as those exhibiting low variability in ratings, have a greater influence on the reconstruction process.

The remainder of the paper is structured as follows, Section \ref{sec:related_works} reviews the related work and discusses group recommendation sec.~\ref{sec:group_recommendation}; matrix completion sec.~\ref{sec:matrix_completion}, and the soft-impute algorithm sec.~\ref{sec:softimpute}. Section ~\ref{sec:proposedApproach} explains the proposed methodology. Section ~\ref{sec:Experiments}, presents the comparative evaluation results. Finally, Section ~\ref{sec:Conclusion} provides a summary of this study, including conclusions and future research directions.

\section{Related Works}\label{sec:related_works}
 In this section, we provide a background to our work, and review the relevant works. Additionally, we present key definitions and concepts. We then provide a brief introduction to the soft-impute algorithm, which inspired our proposed method.

Several approaches in literature have been proposed to address the high-dimensional and sparse nature of recommender system data. One approach integrates dimensionality reduction techniques, using both supervised and unsupervised learning to efficiently address the curse of dimensionality problem, detect user groups, and improve prediction quality \citep{bib23}.

In another research, the authors propose a generalized nonparametric additive model to analyse high-dimensional group testing data to efficiently identify defective items, enabling efficient identification of defective items. The authors utilise the EM algorithm and stochastic gradient descent algorithm to enhance computational efficiency of the group testing procedure \citep{bib24}.
Another author \cite{bib25} utilised a stack ensemble learning framework  for processing high-dimensional group structure data. This work combines the  strength of multiple algorithms and group structure regularization to tackle variable selection in high-dimensional data effectively.

\noindent \cite{bib13} addressed these issues by initially creating matrices based on item-user properties, which were then combined to form the similarity singular value decomposition matrix (SSVD) \citep{bib13}. The authors then developed a context matrix using the contextual weighted property clustering similarity criterion applied to contextual data. The context matrix is then merged with the SSVD matrix to improve recommendation performance and reduce the challenge of cold start and sparse data.
 \cite{bib14}  proposed a cross-domain collaborative filtering method to transfer user-item rating patterns from an auxiliary rating matrix to another domain, alleviating the sparsity of the target domain's rating matrix.
The aggregation for these approaches can be achieved using various techniques such as average without misery \citep{bib4}, average \citep{bib6,bib5}, plurality voting \citep{bib5}, least misery \citep{bib7}, and most pleasure strategies \citep{bib8}.\newline
% 6 Literature review GRS
 Several studies have focused on improving group recommendation and tackling challenges within the group recommendation context.
  \cite{bib9} proposes a latent group model approach  to enhance the accuracy and efficiency of group recommendation. This method involves identifying clusters by applying k-means algorithms on a user-latent factor matrix. The resulting user-factor groups are then combined to create the group profile, which generates group preferences for items using matrix multiplication.  \cite{bib10} focus on SVD-based group recommendation methods, which involve aggregating ratings from group members using various decision strategies. The authors further evaluate the effectiveness of SVD-based aggregation methods, including aggregation profiles and prediction techniques.
Another approach \citep{bib11} combines the clustering algorithm with SVD to enhance the collaborative filtering method, address data sparsity and cold start issues as well.  \cite{bib120} applied the collaborative filtering method on interest subgroups by taking into account user preferences. They then combine recommendation list from all subgroups to form a final recommendation for a large user group. Another paper presents a graph clustering algorithm that leverages pairwise preference data to generate recommendation to user groups \citep{bib26}.\newline

\subsection{Notation}
 We provide a brief summary of notations used throughout the paper, the notations are similar to \citep{bib20,bib33}. Define a matrix $P_{\Omega}(X)$ 
 \[
P_\Omega(X) =
\begin{dcases*}
X_{i,j} \quad if \, (i,j)  \in  \Omega \\
0 \qquad if \, (i,j)  \not\in \Omega
\end{dcases*}
\]
 as the projection of the matrix $X_{m \times n}$ onto the observed entries. Here,  $\Omega \subset \{1, \ldots, m\} \times \{1, \ldots, n\}$ represents the set of indices corresponding to observed entries in $X$. For $(i, j) \in \Omega$, the observed rating is denoted by $X_{ij}$, while $X_{ij} = 0$ for unobserved entries. 
Additionally, from \cite{bib33}
the complementary projection $P\frac{\perp}{\Omega}(X)$ 
is the projection of the matrix $X_{m \times n}$ onto the unobserved entries such that, \break
$P\frac{\perp}{\Omega}(X) + P_\Omega(X) = X$, where, $\Omega \frac{\perp}{ij}=1-\Omega{ij}$ is the complement of $\Omega$.

\subsubsection{Problem Statement: Group Recommendation}\label{sec:group_recommendation}
Let $\mathcal{U} = \{u_i\}_{i=1}^m$ denote the set of $m$ users and $\mathcal{V} = \{v_j\}_{j=1}^n$ represent the set of $n$ items. The interaction between users and items is captured by the partially observed user-item rating matrix $X_{i,j} \in \mathbb{R}^{m \times n}$, where $X_{ij}$ corresponds to the rating provided by user $u_i$ for item $v_j$.

Given a group of users $G$, such that $G \subset \mathcal{U}$, the task is to recommend a list of $k$ items, denoted by $V_G = \{v_j\}_{j=1}^k$, where $V_G \subset \mathcal{V}$. These recommendations are expected to satisfy the preferences of the group of users $G$.

\subsection{Matrix completion and low rank assumption}\label{sec:matrix_completion}

Matrix completion aims to fill in missing entries of a partially observed matrix using its observed entries so that the reconstructed matrix approximates the original complete matrix. Obviously, this is an ill-posed problem and usually requires additional assumptions to the problem. 

To obtain a well-posed problem, it is necessary to make low-rank assumptions on the underlying matrix \citep{bib31}. For instance, the Netflix data matrix, which contains user ratings for movies, is believed be low rank because user's preference for movies is influenced by a small number of latent factors such as movie genre and audience type. 
Low rank techniques are effective for solving problems of matrix completion by exploiting redundant information in interaction matrices and compressing it for more efficient computations. Low-rank models have shown their effectiveness in addressing sparsity problems, imputing missing data, denoising noisy data, and performing feature extraction \citep{bib17,bib16}.

Matrix factorisation techniques are a widely adopted class of algorithms. These methods learn the factors of two or more low-rank matrices by reformulating the problem as an optimization task. However, the rank constraints lead to non-convex objective functions, making the problem inherently NP-hard and computationally intractable. As a result, these problems are typically addressed using approximation techniques that involve various relaxations of the objective functions.

In the following, we present two widely used formulations and algorithms - soft-impute and MF - for addressing the matrix completion problem.

\begin{figure*}[t]
  \centering
  \includegraphics[width=\textwidth]{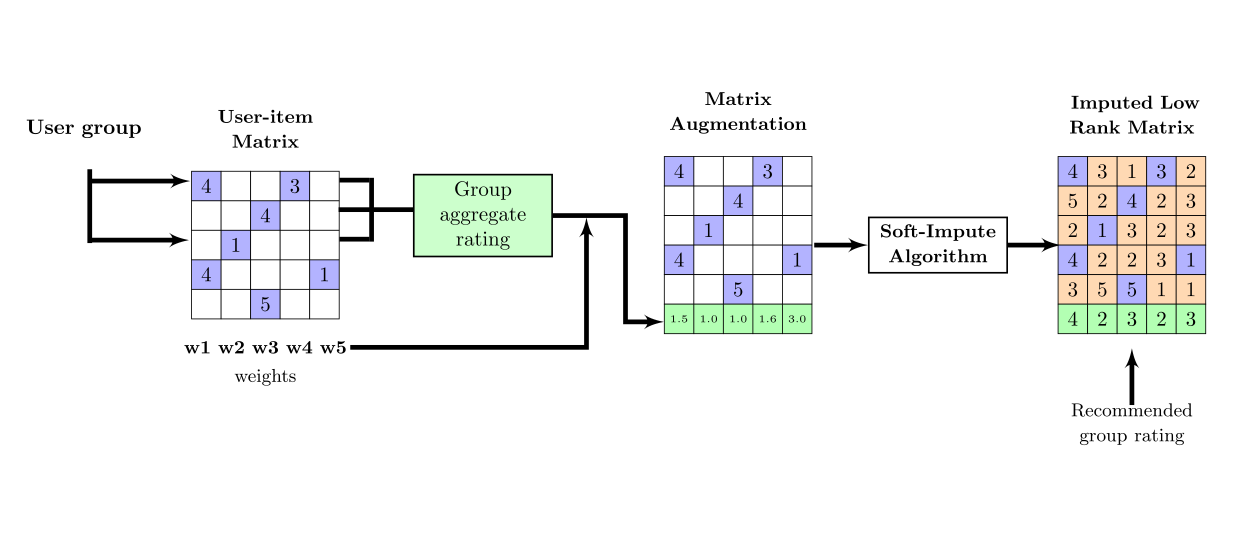}
  \caption{Proposed Methodology}
  \label{fig1}
\end{figure*}

\subsection{Matrix Completion using Soft-Impute Algorithm}\label{sec:softimpute}
A fundamental approach to the matrix completion problem is given in ~\ref{eq:matrixCompletion}, where the rank constraint is relaxed through the nuclear norm regularization term \( ||\mathbf{Z}||_{*} \) \citep{Fazel2004RankMA}. This term, defined as the sum of the singular values of \( \mathbf{Z} \), serves as a convex surrogate for the rank function, enabling a tractable optimization formulation.

\begin{equation}\label{eq:matrixCompletion}
    \min_{\mathbf{Z}} \quad || P_\Omega(\mathbf{X}) - P_\Omega(\mathbf{Z})||^2_F + \lambda ||\mathbf{Z}||_{*}
\end{equation}
where:
\begin{itemize}[noitemsep, topsep=0pt, partopsep=0pt]
 \item \( \mathbf{X} \in \mathbb{R}^{m \times n} \) is the partially observed data matrix;
\item \( \mathbf{Z} \in \mathbb{R}^{m \times n} \) is the completed matrix to be optimized;
\item \( P_\Omega(\cdot) \) is a projection operator that restricts the comparison to the observed entries in \( \Omega \);
\item \( ||\cdot||_F \) denotes the Frobenius norm;
\item \( ||\mathbf{Z}||_{*} \) represents the nuclear norm, defined as the sum of the singular values of \( \mathbf{Z} \);    
\item \( \lambda > 0 \) is a regularization parameter that controls the trade-off between matrix completion accuracy and low-rank structure.  
\end{itemize}
An iterative approach, known as soft-Impute algorithm and based on the SVD of \( \mathbf{Z} \), was proposed by~\cite{bib33}.  
The soft-impute algorithm operates similarly to an expectation-maximization (EM) procedure, where missing values are iteratively updated using the current estimate. The optimization problem is then solved on the completed matrix by applying a soft-thresholded SVD, as described in the following section.

The soft-thresholded SVD is governed by the thresholding operator, defined as follows:  

\begin{equation}\label{eqn4}
 \widetilde{\mathbf{X}} = \mathbf{U} D_{\lambda}(\mathbf{\Sigma}) \mathbf{V}^T 
\end{equation}
where 
\begin{align*}
    D_{\lambda}(\mathbf{\Sigma}) &= \text{diag}[(\sigma_i - \lambda)_{+}], \\
    t_{+} &= \max(0, t).
\end{align*}
  
The operator \( D_{\lambda}(\cdot) \) applies soft-thresholding to the singular values \( \sigma_i \) of \( \mathbf{X} \), shrinking them toward zero. Singular values smaller than \( \lambda \) are set to zero, while those greater than \( \lambda \) are reduced by \( \lambda \).  

By utilizing the soft-thresholding operator in combination with nuclear norm minimization, the soft-impute algorithm iteratively computes a low-rank approximation of \( \mathbf{X} \). 
In addition, \cite{bib33} demonstrated that decreasing the regularization parameter $\lambda$ leads to an increase in the rank of the resulting matrix. Consequently, they suggested initializing each iteration with the solution from the previous step, denoted as $Z_{\lambda_{i-1}}$, where $\lambda_{i-1} > \lambda_i$. The initial value $\lambda_1$ is set as the maximum singular value of the observed matrix $P_\Omega(X)$, i.e.,
\[
\lambda_1 = \sigma_{\max}(P_\Omega(X)),
\]
providing a warm start for the iterative process.

\subsection{Matrix Factorisation and Alternating Least Square}
MF method addresses the matrix completion problem by learning two distinct low-rank matrices of rank \( r \). This rank-constrained formulation is obtained by solving the following optimization problem:

\begin{equation}\label{eq:mmmf}
\resizebox{.60\columnwidth}{!}{$
    \begin{aligned}
        \min_{\mathcal{U}, \mathcal{V}} &\quad || P_\Omega (\mathbf{X}) - P_\Omega (\mathcal{U}\mathcal{V}^T) ||^2_F \\
        &\quad + \lambda (||\mathcal{U}||^2_F + ||\mathcal{V}||^2_F)
    \end{aligned}
$}
\end{equation}

\normalsize

where:
\begin{itemize}
    \item \( \mathcal{U} \) and \( \mathcal{V} \) are latent factor matrices of dimensions \( m \times r \) and \( n \times r \), respectively, where \( m \) represents the number of users and \( n \) represents the number of items.
\end{itemize}
%\vspace{-3.5em}  % Adjust this value as needed
Although the MF formulation is not convex in \( \mathcal{U} \) and \( \mathcal{V} \) simultaneously, it is \textit{bi-convex}. Specifically, for a fixed \( \mathcal{U} \), the objective function is convex in \( \mathcal{V} \), and for a fixed \( \mathcal{V} \), the function is convex in \( \mathcal{U} \). As a result, an alternating least squares technique is commonly employed to minimize \eqref{eq:mmmf}.  
Consider the case where \( \mathcal{U} \) is fixed, and we seek to optimize \( \mathcal{V} \). In this scenario, the problem decomposes into \( n \) separate ridge regression problems, where each column \( \mathbf{x}_j \) of \( P_{\Omega}(\mathbf{X}) \) serves as the response variable, and the \( r \) columns of \( \mathcal{U} \) act as predictors.  
Since some entries in \( \mathbf{x}_j \) are missing, these missing observations are ignored, and the corresponding rows of \( \mathcal{U} \) are removed for the \( j \)-th regression. Consequently, each regression problem operates on a different subset of data, despite all regressions being derived from \( \mathcal{U} \). This formulation ensures that the optimization procedure remains computationally feasible and scalable across large datasets.
However, unlike the soft-impute algorithm, which automatically promotes a low-rank solution through nuclear norm regularization, the rank constraint in the MF approach is imposed through the fixed dimensions of the factorized matrices \( \mathcal{U} \) and \( \mathcal{V} \), which limits the model's flexibility in adapting to the intrinsic rank structure of the data.

\section{Methodology}
\subsection{Proposed Approach}\label{sec:proposedApproach}
In this section, the proposed framework for the group recommender system that forms the basis of our algorithm is illustrated in Fig. ~\ref{fig1} and described below.  Additionally, the details of our algorithm, Group soft-impute SVD (GSI-SVD) are presented in Algorithm ~\ref{Algo2}.

    The first step begins with a user-item matrix, which represents the interactions between users and items. This matrix is a partially observed matrix due to sparse user feedback, where blank squares   
    represent unobserved entries, and blue squares represent observed entries as shown in Fig.~\ref{fig1}.
    
    To incorporate group-based information into the user-item matrix for group recommendation, we compute a group aggregate rating that reflects the combined preferences of the users. This is achieved using a weighted aggregation approach, where each user’s contribution is adjusted by a corresponding weight. To compute the weighted average ratings for group $\mathrm{G}$, only items rated by group members are considered. The average user rating for each item $\mathrm{j}$ in the aggregated group entry is given by  \eqref{eq:group_rate}, while the weights for each item in the aggregated group entry are given by  \eqref{eq:group_weight}, where $\sigma_{G, i}$ is the standard deviation of item $i$ in the group $G$. The weights are based on the proportion of group members who rated the items and the similarity between items rated by the group members, following a similar approach to \cite{bib7}.
    Next, the soft-impute algorithm \citep{bib33} is iteratively applied to the augmented matrix to estimate unknown group preferences by leveraging patterns in the observed data. After the soft-impute algorithm converges, the
result is a low-rank matrix with the missing values predicted. The final step produces a recommended group rating for each item.
   
\vspace{0.3cm}

\begin{algorithm}[t]
\small

\caption{Group Soft-Impute SVD \label{Algo2} }
\begin{algorithmic}[1]
    \Require Input matrix \( X \in \mathbb{R}^{n \times m}\), user group \(i \in G\), number of lambda values \(k\), \(   \lambda_{max}=\sigma_{max}\left(P_\Omega\left(X\right)\right)\), \(\lambda_{min} =1\), convergence threshold \( \epsilon > 0 \)  
    
    \State Randomly initialize $Z^{old}$
    \State \(\lambda_{grid}=[\lambda_{max},...,\ \lambda_{min}]\)
    \State Calculate the group average ratings for observed entries:
    \begin{equation}
r_{G, j} = \frac{1}{|G|} \sum_{i \in G} x_{i, j}, \quad  \text{ where } x_{i,j} > 0. \label{eq:group_rate}
\end{equation}

    \State Compute weights for each item \( j \) as:
 \begin{equation}    
    w_{G, i} = \frac{\# \{ u \in G|u_{u,i} \ne 0 \}}{|G|}  \frac{1}{1 + \sigma_{G, i}}\label{eq:group_weight}
    \end{equation}

    \State Merge weighted group rating \( r_{G,j} \) as an additional row in \( X \):
    \[
    X_{\text{new}} = 
    \begin{bmatrix}
    X \\
    r_{G,j} w_{G,j}
    \end{bmatrix}
    \]

 \For  {\text{each} \, \(\lambda_i \in \lambda_{grid}\)}

\State \resizebox{.95\linewidth}{!}{$
\text{Compute} \: U, D, V^T \gets \mathrm{SVD}\left(P_\Omega(X_{\text{new}}) + P_{\frac{1}{\Omega}}(Z^{\text{old}})\right)
$}

 \State Compute \(D_{\lambda_i}\gets\max{\left(D-\lambda_i,0\right)}\)
 \State $X^{new} \gets U D_{\lambda_i} V^T$
\If{$\frac{ \left\|Z^{new}-Z^{old}\right\|_F^2}{\left\|Z^{old} \right\|_F^2} < \epsilon$}
\State break
\EndIf
\State $Z^{old} \gets Z^{new}$
\EndFor
\State \(Z_{\lambda_\kappa}\gets Z^{old}\)
\State \(r_{G_j}\gets Z_{\lambda_\kappa}[-1]\)
\State Output: \text{Final reconstructed matrix} \(Z_{\lambda_\kappa}\) , \text{Predicted group} rating \(r_{G_j}\)
 
\end{algorithmic}
\end{algorithm}
%GRS
\normalsize

\subsection{Computational Complexity and Convergence}

The proposed GSI-SVD algorithm achieves efficient computation by combining iterative singular value thresholding with a dynamic low-rank matrix reconstruction strategy that retains only the non-zero singular components at each step.

At each iteration, the algorithm performs soft-thresholding on the singular values. For an input matrix \( X \in \mathbb{R}^{n \times m} \), the SVD is applied to the matrix
\[
Y = P_\Omega(X) + P_{\Omega^\perp}(Z_{\text{old}}),
\]
which has a worst-case computational cost of \( \mathcal{O}(nm^2) \). However, after thresholding, only a small number of singular components \( r \ll \min(n, m) \) are typically retained. Reconstructing the matrix using only these \( r \) components reduces the per-iteration cost to:
\[
\mathcal{O}(nmr).
\]
Additionally, applying the mask over the observed entries \( \Omega \) introduces a further cost of \( \mathcal{O}(|\Omega|) \). Therefore, the total computational cost per iteration becomes:
\[
\mathcal{O}(|\Omega| + nmr).
\]

Following the convergence analysis in \citep{bib33}, the algorithm guarantees that the relative error between consecutive iterates,
\begin{equation}
\frac{\left\|Z^{(k+1)} - Z^{(k)}\right\|_F}{\left\|Z^{(k)}\right\|_F},
\label{eq:relative_error}
\end{equation}
tends to zero as the number of iterations \( k \) increases. Under mild assumptions, this results in geometric convergence (See Appendix  for the convergence proof), with the number of iterations required to reach a desired accuracy \( \epsilon \) scaling as:
\[
k = \mathcal{O}(\log(\epsilon^{-1})).
\]

This makes the GSI-SVD algorithm well-suited for large-scale, sparse matrix completion in group recommendation scenarios.

\subsection{Baseline}
We evaluate our proposed method by comparing it against two established techniques: the weighted before factorisation method and the after factorisation method \citep{bib7}. Unlike the soft-impute algorithm, these group recommendation methods are derived MF approaches.  

The MF-based group recommendation framework is extended in two ways: 
\begin{enumerate}
    \item Ratings for a selected group of users are first aggregated using an aggregation function \( \mathbf{h} \) to construct a pseudo-group user \( u_g \), before applying factorization to the rating matrix. 
    \item Ratings are computed after aggregating predictions derived from the latent factor matrices.  
\end{enumerate} 
These two techniques serve as strong baselines for evaluating group recommendation systems within collaborative filtering frameworks. By effectively capturing and integrating user preferences, they provide a robust benchmark for assessing the performance of the proposed approach.  

In the following sections, we present a detailed description of these methods.
\subsubsection*{After Factorisation Method}
The after factorisation (AF) approach proposed in the literature \citep{bib7} recommends items to a group of users by leveraging the latent factors learned through MF. Initially, MF model is trained to obtain latent factor representations for both users and items, effectively capturing their underlying preferences and characteristics.  

To model group preferences, the individual latent factors of the group members are aggregated into a group profile \( \mathbf{u}_G \) and item profile \( \mathbf{v}_G \), respectively, using a predefined aggregation function \( \mathbf{h} \).  

Commonly used aggregation functions include the average, weighted average, minimum aggregation and maximum aggregation. 
\subsubsection*{Weighted Before Factorization Method }
The weighted before factorization (WBF) approach \cite{bib7} focuses on computing a weighted aggregation of the individual preferences in the user rating matrix before applying MF. By emphasizing the importance of certain entries prior to factorization, this method constructs a collective preference profile for the group while ensuring a degree of fairness among its members.

For a group \( G \) with a set of items, weights \( w_{G,j} \) are defined as in equation \ref{eq:group_weight}
The group user  profile \( \mathbf{u}_G \) is computed by solving the objective function described in equation~\ref{eq:mmmf}.

\section{Experiments}\label{sec:Experiments}
\subsection{Datasets}
The experiments were performed on the well-known real-world Goodbooks dataset \citep{bib39}, Movielens dataset \cite{movielens} and a synthetic dataset.

\begin{table}[h]
\caption{Goodbooks dataset }\label{Table1}%
\begin{tabularx}{.5\textwidth}{@{}l|l}
\toprule
Dataset & Detail  \\
\midrule
\#number of users & 53K  \\
\hline
\#number of books & 10K \\
\hline
\#number of ratings      & 6 million \\
\botrule
\end{tabularx}
\end{table}

The Goodbooks dataset (Table \ref{Table1}) contains ratings on a scale from 1 to 5, provided by various users for different books. For computational purposes, we select a sample of 2000 users and 200 books from the dataset. The data has a sparsity of 90\%, indicating the proportion of unobserved entries.
\begin{table}[h]
\caption{Movielens dataset }\label{Table2}%
\begin{tabularx}{.5\textwidth}{@{}l|l}
\toprule
Dataset & Detail  \\
\midrule
\#number of users & 943  \\
\hline
\#number of movies & 1682 \\
\hline
\#number of ratings      & 100K \\
\botrule
\end{tabularx}
\end{table}
\vspace{0.3cm}
The Movielens dataset contains 100,000 ratings provided by 943 users for 1,682 items. Each user has rated at least 20 movies. The ratings span a scale from 1 to 5 and have a sparsity of 86\%. For this study, we selected a sample of 943 users and 500 items from the data set. 
For the matrix completion experiments, we first use the KNN imputer algorithm to estimate the missing entries. We then create a partially observed matrix by randomly selecting 75\% of the observed ratings for training and masking the remaining 25\% for testing.

For the synthetic data, we generated a synthetic user-item rating dataset. The dataset consisted of 2,000 users and 200 items, with ratings drawn from a bounded normal distribution centred around a mean rating of 3.5 and a standard deviation of 0.65. The ratings were clipped to fall within the appropriate range, ensuring alignment with typical rating scales used in real-world recommender systems. 
To introduce sparsity, a binary mask was applied, with 25\% of entries observed and 75\% missing entries, simulating real-world missing data scenarios.

\subsection{Experimental Results}

In the first phase of the experiment, we employ the soft-impute algorithm to demonstrate the effectiveness of the nuclear norm as a convex surrogate for recovering low-rank matrices. 
\begin{figure}[t]
\centering

  \noindent\includegraphics[width=\columnwidth]{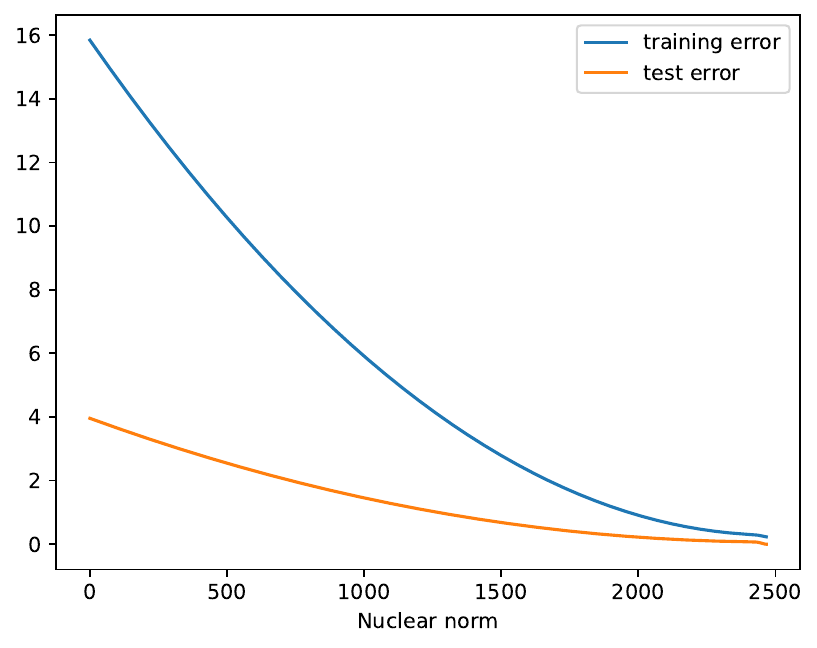}
  \captionof{figure}{Test error-Training error vs Nuclear norm.}
  \label{fig:2}
\end{figure}

All matrix completion computations are implemented using the PyTorch library, which provides efficient support for tensor operations and GPU acceleration.

Soft-impute iteratively fills in the missing entries in the partially observed matrix \( X \) using the current iterate \( Y_t \), evaluated on a grid of 10 values of the regularization parameter \( \lambda \). Convergence is achieved when the relative change in Frobenius norm between successive iterates falls below \( 10^{-5} \).

To reduce computational complexity and align with the efficient \textit{soft-impute} framework proposed by \citet{bib33}, we modify the standard SVD-based reconstruction in our GSI-SVD implementation. Rather than reconstructing the entire matrix using all singular vectors and a full-size diagonal matrix, we retain only the singular values \( \sigma'_i > 0 \) that remain positive after soft-thresholding. The low-rank matrix is then reconstructed using only the reduced set of singular vectors:
\[
Z_{\text{new}} = U_r \cdot \Sigma_r \cdot V_r^\top,
\]
where \( U_r \in \mathbb{R}^{n \times r} \), \( \Sigma_r \in \mathbb{R}^{r \times r} \), and \( V_r \in \mathbb{R}^{m \times r} \) correspond to the singular vectors and values associated with the retained \( r \) non-zero components.

\begin{figure}[h]
\centering
\includegraphics[width=\columnwidth]{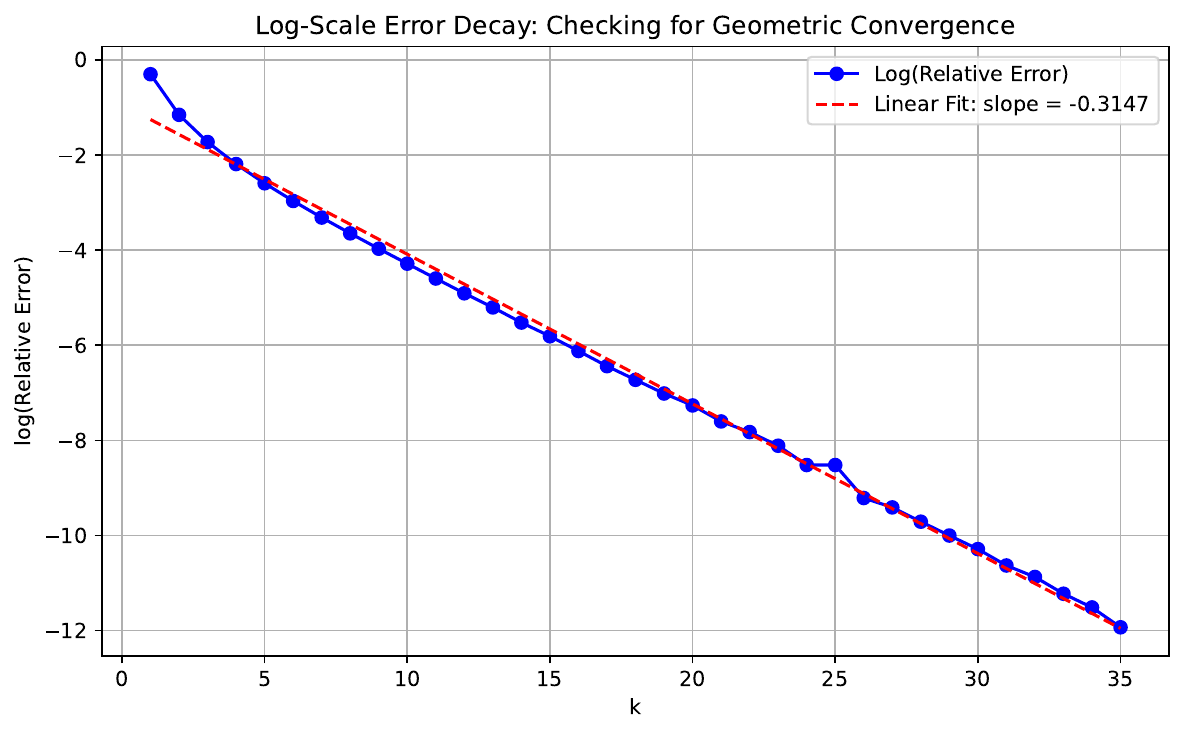}
  \captionof{figure}{Log-scale plot of the relative error across iterations.}
  \label{fig:4}
\end{figure}

\begin{figure*}[!t]
    \centering
    \subfigure (a) Goodbooks dataset{
        \includegraphics[height=5.5cm,width=0.9\textwidth]{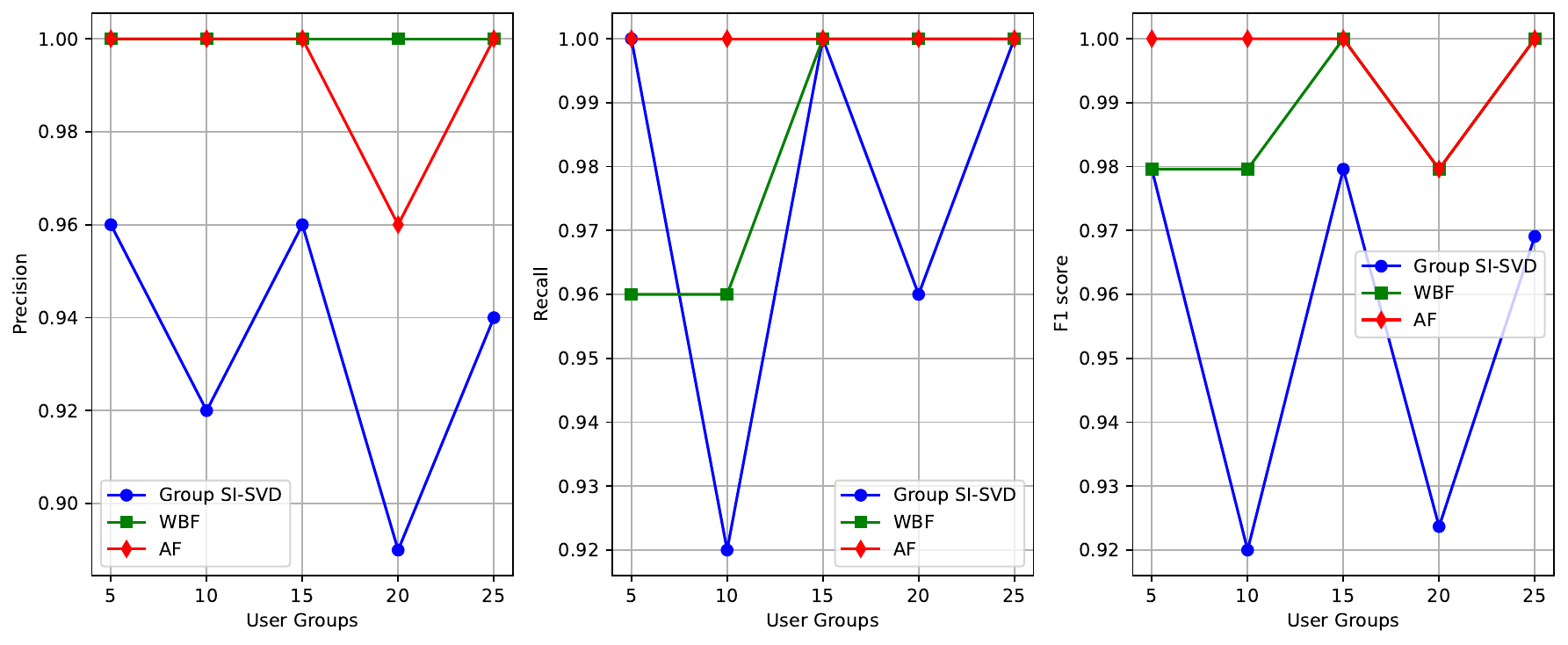}
        \label{fig:sub1}
    }

    \subfigure (b) Movielens dataset{
        \includegraphics[height=5.5cm,width=0.9\textwidth]{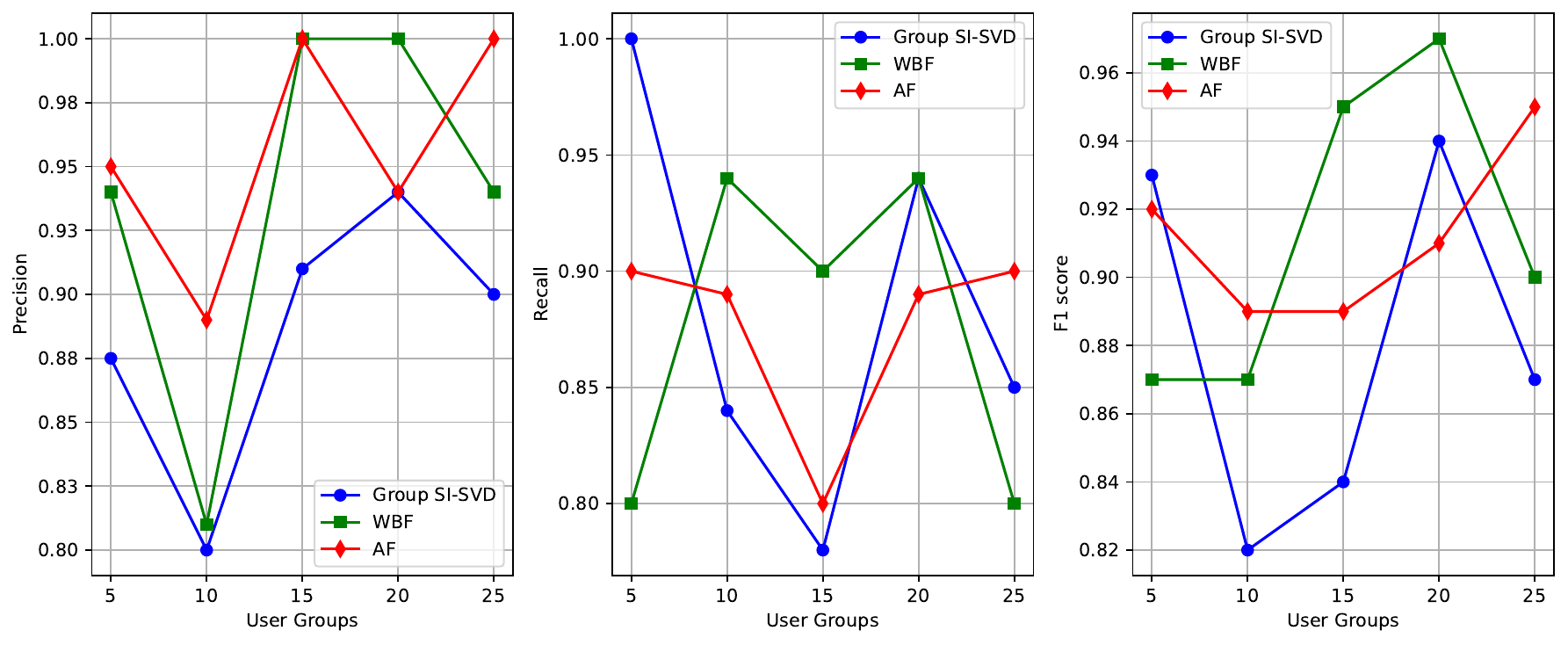}
        \label{fig:sub2}
    }

    \subfigure (c) Synthetic dataset{
        \includegraphics[height=5.5cm,width=0.9\textwidth]{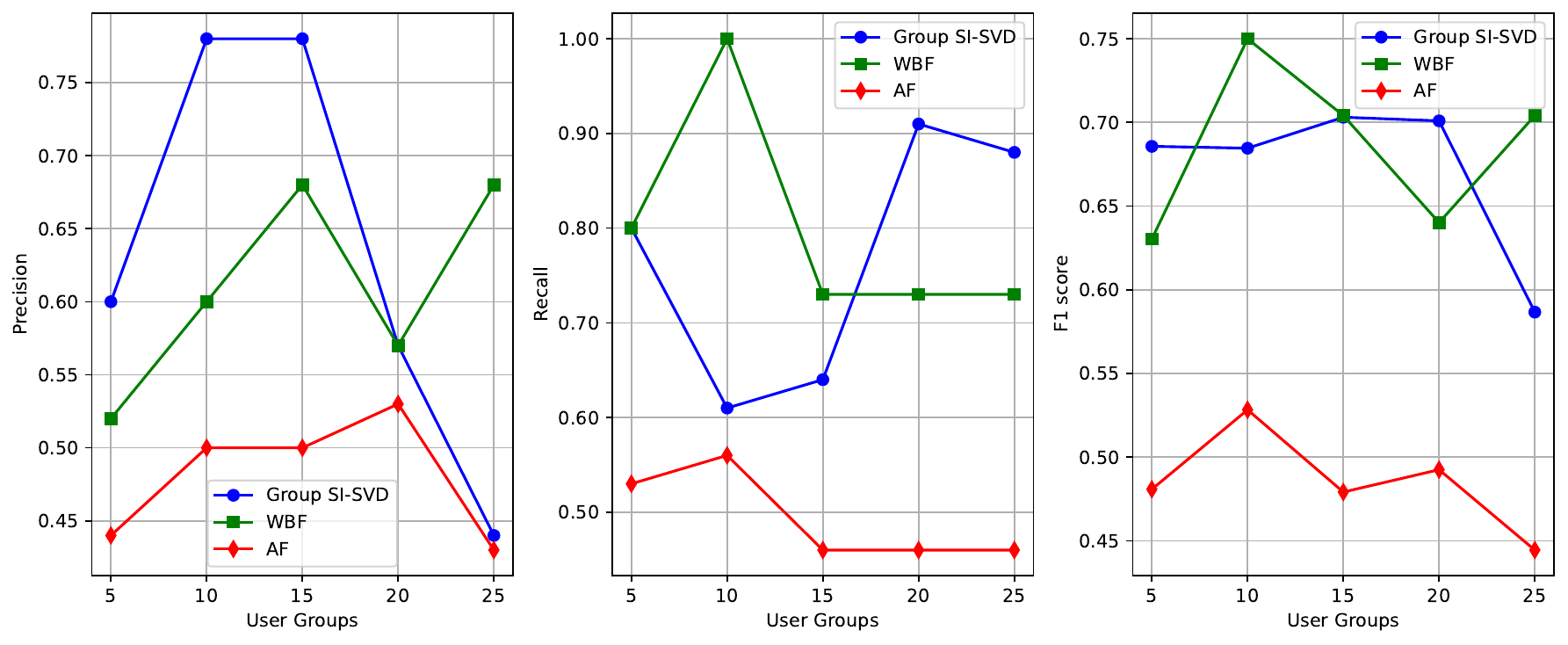}
        \label{fig:sub3}
    }

    \caption{Performance comparison of GSI-SVD, AF, and WBF for different user group sizes across (a) Goodbooks, (b) Movielens, and (c) Synthetic datasets.}
    \label{fig:3}
\end{figure*}
For performance evaluation, 
\begin{itemize}[noitemsep, topsep=6pt, partopsep=0pt]
\item We use the mean squared error(MSE) for training error and test error defined as:
\begin{equation*}\vspace{-3mm}
   \text{Training error}=\frac{P_{\Omega}(X-Y)^2}{|\Omega|}
\end{equation*}\vspace{-3mm}
\begin{equation*}
  \text{Test error}=\frac{P_\frac{1}{\Omega}(X-Y)^2}{|\frac{1}{\Omega}|} 
\end{equation*}
\end{itemize}

Figure \ref{fig:2} displays plots of the training and test error for the GSI-SVD algorithm as a function of nuclear norm over the grid of $\lambda$ $\in$ [$\lambda_{max}-|n|$,0], where $\lambda_{max}$ corresponds to the largest singular value obtained from the SVD of $P_\Omega(X)$ and $|n|$ is the cardinality of the items in $X$.

The results in Figure \ref{fig:2} show the performance of the algorithm for datasets with 75\% missing entries. The results also indicate that soft-impute effectively recovers the matrix with high nuclear norm.

 \begin{table*}[!t]
\centering
\footnotesize  % Slightly larger than %\scriptsize
\setlength{\tabcolsep}{13.0pt}  % Increased spacing between columns
\caption{Recovered ranks for WBF, AF, and GSI-SVD at varying $\lambda$ values across datasets.}
\begin{tabular}{c|ccc|ccc|ccc}
\toprule
$\lambda$ & \multicolumn{3}{c|}{GoodBooks} & \multicolumn{3}{c|}{Movielens} & \multicolumn{3}{c}{Synthetic} \\
\cmidrule(lr){2-4} \cmidrule(lr){5-7} \cmidrule(lr){8-10}
 & WBF & AF & GSI & WBF & AF & GSI & WBF & AF & GSI \\
\midrule
0.001 & 199 & 199 & 200 & 498 & 497 & 491 &  44 &  44 & 200 \\
0.01  & 200 & 199 & 176 & 498 & 498 & 270 & 199 & 199 & 182 \\
0.1   & 197 & 196 & 174 & 496 & 496 & 255 & 198 & 196 & 182 \\
1     & 198 & 197 & 166 & 497 & 498 & 237 & 198 & 198 & 175 \\
10    & 198 & 199 &  72 & 499 & 498 &  82 & 198 & 199 & 111 \\
\bottomrule
\end{tabular}

\label{Table4}
\end{table*}

To evaluate the performance of the group recommendation model, we compute precision, recall and F1 score that compare the aggregated individual-rating predictions $r(G,i)$ with the aggregated group predictions  $\hat{r}(G,i)$, following a similar approach to prior work \cite{bib40}.
\bigskip
 \begin{equation}\label{eq10}
 r(G,i)= \frac{\sum_{u \in G} x(i,j)}{|G|}
 \end{equation}
 \bigskip
 \begin{equation}\label{eq11}
 \hat{r}(G,i)= \frac{\sum_{u \in G} \hat{x}(i,j)}{|G|}
 \end{equation}

 \bigskip
 
\noindent \textbf{Precision} \citep{bib45} is the proportion of recommended items that are relevant, calculated as the ratio of relevant recommended items to the total number of recommended items.
\begin{equation}\label{eq12}
\mathrm{precision = \frac{TP}{TP+FP}}
\end{equation}

\noindent \textbf{Recall} \cite{bib45} is the ratio of the number of relevant recommended items to the total number of relevant items in the dataset. 
\begin{equation}\label{eq12}
\mathrm{recall = \frac{TP}{TP+FN}}
\end{equation}
where TP, FP, and FN denote true positives, false positives, and false negatives, respectively.

\bigskip
\noindent \textbf{F1 score} \citep{bib_f1} is the harmonic mean of precision and recall given by:
\begin{equation}\label{eq13}
\text{F1 score} = 2 \cdot \frac{\text{precision} \cdot \text{recall}}{\text{precision} + \text{recall}}
\end{equation}
These metrics are typically evaluated at a predefined recommendation list size $k$. The results show that the proposed framework is effective in capturing the preferences of the group members and providing accurate group-level recommendations.

To evaluate the proposed method (GSI-SVD), we compared it against the baselines (WBF and AF approach). The comparison was based on precision, recall and F1 score for the top K recommended items. Figure~\ref{fig:3} present comparisons for different user groups of sizes 5, 10, 15, 20 and 25 against $\mathrm{k}$ number of item recommendation, where ($\mathrm{k}$=20) for this experiment.

For the Goodbooks data, the experimental results show that the WBF and AF methods outperform the proposed method in terms of precision across all user groups. In general, AF and WBF demonstrate better precision, indicating their superior ability to recommend relevant books with fewer false positives. In terms of recall, GSI-SVD outperforms WBF for user group 5, but shows similar performance to baseline methods for group sizes 15 and 25, as shown in Figure ~\ref{fig:3}a .
AF maintains a perfect F1 score in all groups, indicating high precision and recall. In contrast, GSI-SVD exhibits the highest variation in F1 score, peaking at around $\sim$0.98 for group 15 before dropping to approximately $\sim$0.92 for group 20.
For the Movielens dataset, AF and WBF outperform GSI-SVD in terms of recommendation accuracy for some user groups, although the differences are relatively small. GSI-SVD shows lower precision for smaller group sizes of 5 and 10, but its performance improves for larger group sizes. While GSI-SVD achieves a higher recall than WBF and AF for user group 5, the latter two methods outperform it for user groups 10 and 15. Additionally, GSI-SVD has the highest F1 score for user group 5, but is outperformed by WBF and AF in the mid-sized groups as shown Figure ~\ref{fig:3}b .

For the Synthetic data set, GSI-SVD maintains a consistently higher precision than the other methods broadly across different group sizes as shown in Figure~\ref{fig:3}c. Both GSI-SVD and WBF outperform AF, which shows the lowest precision across all groups. In terms of recall, GSI-SVD  demonstrates improved performance as the group size increases. Conversely, AF consistently shows low recall, suggesting it is missing many relevant recommendations. Both GSI-SVD and WBF demonstrate more competitive performance in terms of F1 score compared to AF. This suggests that GSI-SVD and WBF are the most effective models for achieving high recall, while AF is the least effective for this data set.

We also compare GSI-SVD, WBF, and AF in terms of rank recovery across different $\lambda$ values. WBF and AF seem to recover approximately the full rank \(\sim 200\) and \(\sim 497\) for Movielens across all $\lambda$ values whereas GSI-SVD shows a strong dependence on $\lambda$, with ranks decreasing significantly as $\lambda$ increases as shown in Table~\ref{Table4}. 

To empirically verify the convergence behavior of the proposed method, we analyze the decay of the relative error \ref{eq:relative_error} between successive iterates \( k \), as shown in Figure~\ref{fig:4}. Specifically, we plot the logarithm of the relative error,
 against the iteration count. The observed near-linear trend with a negative slope on the log-scale plot indicates exponential decay of the error, thereby validating the geometric convergence behavior of the GSI SVD algorithm. 
Therefore, the method used in this paper significantly enhances the quality of recommendations in terms of precision and recall for small, mid size and large user groups, as shown in Figure \ref{fig:3}. Furthermore, the incorporating weights to group recommendation significantly enhances the quality of recommendations in terms of precision and recall showing that the predictions are aligned with the group's collective preferences.

\section{Conclusion}\label{sec:Conclusion}
In this paper,  a novel recommendation method for collaborative filtering that imputes missing entries in a matrix and offers recommendation to a given user group leveraging soft-impute algorithm. The results show that the soft-impute algorithm is effective in recovering the underlying low-rank structure of the partially observed matrix and providing group recommendation that satisfy the preferences of the  small size, mid-size and large number of group members. The results indicate that the recommendation process improves both precision and recall across varying user group sizes (5, 10, 15 20 and 25) and different numbers of item recommendations.
In future studies, we aim to investigate the performance of the proposed GSI-SVD algorithm for even larger group sizes, evaluating its effectiveness and scalability. Additionally, we plan to develop optimization techniques to address the computational challenges of large group recommendations while maintaining accuracy.

\bibliographystyle{apalike}

\section*{Appendix : Convergence}\label{appendix}

While the theoretical convergence of the original Soft-Impute algorithm is generally sublinear \citep{bib33},  modifications introduced in Group Soft-Impute SVD, combined with empirical observations, support the assumption of linear(geometric) convergence rate.
The soft-thresholded SVD operator, which is fundamental to the algorithm, is known to be non-expansive in the Frobenius norm \citep{bib33}:
\[
\|\text{D}_\lambda(A) - \text{D}_\lambda(B)\|_F \leq \|A - B\|_F.
\]
However, when the iterates approach a stable low-rank solution manifold, soft-thresholded SVD can exhibit \textit{contractive} behavior, satisfying:
\[
\|Z^{(k+1)} - Z^*\|_F \leq \rho \|Z^{(k)} - Z^*\|_F, \quad \text{for some } 0 < \rho < 1,
\]
where \( Z^* \) denotes the optimal low-rank solution. This contraction behaviour implies that the iterates converge exponentially toward the solution. Prior studies on low-rank matrix projection and proximal methods confirm that such contraction is expected when the iterates remain within a stable low-rank subspace, thus supporting the empirical linear convergence observed in our results.

\subsection*{Convergence Derivation}
Assume that the iterates $Z_k$ of the Group Soft-Impute SVD algorithm converge geometrically to the optimal solution $Z_\infty$, that is,
\[
\|Z_k - Z_\infty\|_F \leq \rho^k \|Z_0 - Z_\infty\|_F, \quad \text{for some } 0 < \rho < 1.
\]

We aim to find the number of iterations $k$ such that the reconstruction error is bounded by a desired accuracy $\epsilon > 0$:
\[
\|Z_k - Z_\infty\|_F \leq \epsilon.
\]

From the geometric decay:
\[
\rho^k \|Z_0 - Z_\infty\|_F \leq \epsilon.
\]

Dividing both sides and taking the logarithm:
\[
k \log(\rho) \leq \log\left(\frac{\epsilon}{\|Z_0 - Z_\infty\|_F}\right),
\]

which gives:

\[
k \geq \frac{ \log\left( \frac{\|Z_0 - Z_\infty\|_F}{\epsilon} \right) }{ -\log(\rho) } = \mathcal{O}(\log(1/\epsilon)).
\]
Thus, the number of iterations grows logarithmically with the inverse of the target accuracy, consistent with the geometric convergence we observe in practice.
\end{document}